\documentclass[a4paper]{jpconf}
\usepackage{graphicx}
\begin{document}
\title{Problems and paradoxes of the Lifshitz theory}

\author{G. L. Klimchitskaya\footnote{On leave from 
North-West Technical University, St.Petersburg, Russia}}

\address{Center of Theoretical Studies and Institute for Theoretical
Physics, Leipzig University, {\protect \\}
Vor dem Hospitaltore 1, 100920,
D-04009, Leipzig, Germany}

\ead{galina.klimchitskaya@itp.uni-leipzig.de}

\begin{abstract}
The problems and paradoxes of the Lifshitz theory in application to
real dielectric and semiconductor materials are reviewed. It is shown 
that the inclusion of drift current of conduction electrons into the 
model of dielectric response results in contradictions with both 
thermodynamics and experimental data of different experimental 
groups. Physical reasons why the problems and paradoxes arise are 
analyzed and found to be connected with the violation of basic 
applicability condition of the Lifshitz theory, the thermal 
equilibrium. A recent alternative approach to the resolution of the 
problems based on the inclusion of screening effects and diffusion 
current is considered and demonstrated to be thermodynamically and 
experimentally inconsistent. It is argued that the inclusion of the 
diffusion current leads to an even deeper violation of thermal 
equilibrium. Phenomenologically, the Lifshitz theory with role of 
drift and diffusion currents neglected is shown to be free of 
problems and in agreement with both thermodynamics and all available
experimental data.
\end{abstract}

\section{Introduction}

It is common knowledge that the Lifshitz theory gives a description of the
van der Waals and Casimir forces between two plane parallel plates
(semispaces) in terms of the frequency-dependent dielectric permittivities
of the plate materials $\varepsilon(\omega)$ [1--3].
Starting in 2000, problems and paradoxes with the 
application of the Lifshitz theory to real
materials (at first to metals [4] and then to dielectrics [5]) were
recognized.
Specifically, for metals described by the Drude dielectric function a
large thermal effect was predicted at short separations [4] in
qualitative disagreement with the case of ideal metals.
This prediction was excluded experimentally at a 99.9\% confidence 
level [6,7]. At large separations, for the Drude metals one half of the
Casimir force acting between ideal metals was obtained [4]. 
The same result was obtained in [8,9] by modelling a metal with a set
of nonrelativistic hard-core or point-like particles confined by walls.
In [10] it
was shown that the substitution of the Drude dielectric function into
the Lifshitz formulas results in a violation of the third law of
thermodynamics for metals with perfect crystal lattices. 
In [11] it was demonstrated that for metals with impurities the Nernst
heat theorem is satisfied. This, however, does not solve the problem
because in accordance with quantum statistical physics the Nernst theorem
must be satisfied for metals with perfect crystal lattices as well.
The situation is
even more dramatic for dielectrics, the materials for which the
Lifshitz theory was originally formulated. It was shown that the inclusion
of the dc conductivity arising in all dielectrics at nonzero temperature
radically changes the theoretical predictions and leads to disagreement
between the Lifshitz theory and thermodynamics [5,12,13].

In this paper we review recent paradoxial applications of the Lifshitz
theory to real materials (mostly to dielectrics) and analyze the
physical reasons of why the paradoxes appear. We show that difficulties arise
each time when the drift or diffusion currents are included in the model
of the dielectric response. In this connection we discuss recent
experiments with semiconductors [14,15] and dielectric [16] test bodies
and show that the measurement results are in disagreement with theory taking
the dc conductivity into account.
Recent theoretical approaches attempting to solve problems of the Lifshitz
theory by accounting for screening effects and diffusion currents [17,18]
are demonstrated to be in contradiction with both thermodynamic
and experimental data.

Our analysis leads to the conclusion that the roots of the problems are
in the violation of the applicability conditions of the Lifshitz theory
when the role of free charge carriers is included in the model of the
dielectric response. We argue that in this case one violates the 
condition of thermal
equilibrium by admitting a unidirectional flux of heat from the Casimir
plates to the heat reservoir.

In Sec.~2 we discuss the consistency of the Lifshitz theory with
thermodynamics. Sec.~3 is devoted to the experimental tests of the
Lifshitz theory under different assumptions about the model of the dielectric
response. Sec.~4 contains the formulation of a phenomenological approach
of how to apply theory in agreement with both thermodynamics and the
experimental data. In Sec.~5 we provide thermodynamic and
experimental tests for the proposed modification of the reflection
coefficients due to screening effects and diffusion current. 
Our discussion and conclusions
are presented in Sec.~6.

\section{Consistency of the Lifshitz theory with thermodynamics}

The main result of the Lifshitz theory is the expression for the free
energy of dispersion interaction per unit area of two semispaces
at a separation distance $a$, at temperature $T$ in thermal equilibrium with
an environment [1--3]
\begin{eqnarray}
&&
{\cal F}(a,T)=\frac{k_BT}{2\pi}\sum_{l=0}^{\infty}
{\vphantom{\sum}}^{\prime}\int_{0}^{\infty}k_{\bot}dk_{\bot}
\left\{\ln\left[1-r_{\rm TM}^{(1)}({\rm i}\xi_l,k_{\bot})
r_{\rm TM}^{(2)}({\rm i}\xi_l,k_{\bot})\,{\rm e}^{-2aq_l}\right]\right.
\nonumber \\
&&\phantom{{\cal F}(a,T)}
+\left. 
\ln\left[1-r_{\rm TE}^{(1)}({\rm i}\xi_l,k_{\bot})
r_{\rm TE}^{(2)}({\rm i}\xi_l,k_{\bot})\,{\rm e}^{-2aq_l}\right]\right\}.
\label{eq1}
\end{eqnarray}
\noindent
Here, $k_{\bot}$ is the magnitude of a wave vector in the plane of
boundary planes (perpendicular to the $z$-axis), prime adds a multiple 1/2 
to the term  $l=0$, $\xi_l=2\pi k_BTl/\hbar$
($l=0,\,1,\,2,\,\ldots$) are the Matsubara frequencies, $k_B$ is the
Boltzmann constant, and the Fresnel reflection coefficients along
the imaginary frequency axis are given by ($n=1,\,2$)
\begin{equation}
r_{\rm TM}^{(n)}({\rm i}\xi_l,k_{\bot})=
\frac{\varepsilon_l^{(n)}q_l-k_l^{(n)}}{\varepsilon_l^{(n)}q_l+k_l^{(n)}},
\qquad
r_{\rm TE}^{(n)}({\rm i}\xi_l,k_{\bot})=
\frac{q_l-k_l^{(n)}}{q_l+k_l^{(n)}},
\label{eq2}
\end{equation}
\noindent
where
\begin{equation}
q_l^2=k_{\bot}^2+\frac{\xi_l^2}{c^2}, \qquad
{k_l^{(n)}}^2=k_{\bot}^2+\varepsilon_l^{(n)}\frac{\xi_l^2}{c^2}, \qquad
\varepsilon_l^{(n)}\equiv\varepsilon^{(n)}({\rm i}\xi_l).
\label{eq2a}
\end{equation}

The free energy of atom-wall interaction, where $\varepsilon_l^{(1)}$
is the dielectric permittivity of wall material, can be expressed as
\begin{eqnarray}
&&
{\cal F}^{A}(a,T)=-k_BT\sum_{l=0}^{\infty}
{\vphantom{\sum}}^{\prime}\alpha({\rm i}\xi_l)
\int_{0}^{\infty}k_{\bot}dk_{\bot}q_l{\rm e}^{-2aq_l}
\label{eq3} \\
&&\phantom{{\cal F}^{A}(a,T)}
\times 
\left\{2r_{\rm TM}^{(1)}({\rm i}\xi_l,k_{\bot})
-\frac{\xi_l^2}{q_l^2c^2}
\left[r_{\rm TM}^{(1)}({\rm i}\xi_l,k_{\bot})+
r_{\rm TE}^{(1)}({\rm i}\xi_l,k_{\bot})\right]\right\},
\nonumber
\end{eqnarray}
\noindent
where $\alpha({\rm i}\xi_l)$ is the dynamic atomic polarizability.

For dielectrics, the dielectric permittivity along the
imaginary frequency axis is commonly presented in the form [19]
\begin{equation}
\varepsilon({\rm i}\xi)=1+\sum_{j}
\frac{g_j}{\omega_j^2+\xi^2+\gamma_j\xi}\approx
1+\sum_{j}\frac{g_j}{\omega_j^2+\xi^2},
\label{eq4}
\end{equation}
\noindent
where $\omega_j\neq 0$ are the oscillator frequencies, $g_j$ are the
oscillator strength and $\gamma_j$ are the relaxation parameters.
This representation assumes some finite value of the static dielectric
permittivity
\begin{equation}
\varepsilon_0\equiv 
\varepsilon(0)=1+\sum_{j}
\frac{g_j}{\omega_j^2}<\infty.
\label{eq5}
\end{equation}
\noindent
Using this model one can find the asymptotic behavior of the free energy of
the Casimir interaction between two similar dielectric plates with
$\varepsilon^{(1)}=\varepsilon^{(2)}=\varepsilon$ at low temperature
$T\ll T_{\rm eff}\equiv \hbar c/(2ak_B)$ [5]
\begin{equation}
{\cal F}(a,T)\approx E(a)-\frac{\hbar c}{32a^3}\zeta(3)r_0^2
(\varepsilon_0+1)\left(\frac{T}{T_{\rm eff}}\right)^3.
\label{eq6}
\end{equation}
\noindent 
Here, $E(a)$ is the Casimir energy at zero temperature, $\zeta(z)$ is the
Riemann zeta function, and $r_0\equiv(\varepsilon_0-1)/(\varepsilon_0+1)$
is the value of the TM reflection coefficient at zero frequency. 
A similar asymptotic expression can be obtained from (\ref{eq3}) for the
free energy of the atom-wall interaction [20]
\begin{equation}
{\cal F}^{A}(a,T)\approx E^{A}(a)-\frac{\hbar c\pi^3}{240a^4}
\alpha(0)C_{D}(\varepsilon_0)
\left(\frac{T}{T_{\rm eff}}\right)^4,
\label{eq7}
\end{equation}
\noindent 
where $C_{D}(\varepsilon_0)$ is some constant found in [20].

Equations (\ref{eq6}) and (\ref{eq7}) allow one to find the asymptotic
behavior of the Casimir entropy in the plate-plate and atom-plate
configurations
\begin{equation}
S(a,T)=-\frac{\partial{\cal F}(a,T)}{\partial T},
\qquad
S^{A}(a,T)=-\frac{\partial{\cal F}^{A}(a,T)}{\partial T}
\label{eq8}
\end{equation}
\noindent 
at low temperature. Thus, from (\ref{eq6}) and (\ref{eq7}) one obtains
\begin{eqnarray}
&&
S(a,T)\approx \frac{3k_B}{16\pi a^2}\zeta(3)r_0^2
(\varepsilon_0+1)\left(\frac{T}{T_{\rm eff}}\right)^2,
\nonumber \\
&&
S^{A}(a,T)\approx \frac{\pi^3k_B}{30a^3}
\alpha(0)C_{D}(\varepsilon_0)
\left(\frac{T}{T_{\rm eff}}\right)^3.
\label{eq9}
\end{eqnarray}
\noindent 
As is seen from (\ref{eq9}), in both cases the Casimir entropy goes to
zero when $T$ vanishes, i.e., the Lifshitz theory with the dielectric
permittivity (\ref{eq4}) satisfies the Nernst heat theorem.

The model of the dielectric response (\ref{eq4}) is, however, not exact.
It does not take into account that at $T\neq 0$ all dielectric
materials possess small but physically real dc conductivity $\sigma(T)$.
For all dielectric materials, the general behavior of $\sigma(T)$ with
vanishing temperature is given by
\begin{equation}
\sigma(T)\sim\exp\left(-\frac{C}{k_BT}\right)\to 0\quad
\mbox{when}{\ }T\to 0,
\label{eq10}
\end{equation}
\noindent
where the physical meaning of the coefficient $C$ is different for
different classes of dielectrics [21,22]. The dc conductivity of
dielectrics is usually taken into account by means of the Drude-like
term (see, e.g., [23])
\begin{equation}
\tilde\varepsilon({\rm i}\xi, T)=\varepsilon({\rm i}\xi)+
\frac{4\pi\sigma(T)}{\xi}.
\label{eq11}
\end{equation}

It would be reasonable to expect that the substitution of the dielectric
permittivity (\ref{eq11}), instead of (\ref{eq4}), into the Lifshitz formulas
(\ref{eq1}) and (\ref{eq3}) leads to only minor corrections to the
asymptotic results (\ref{eq6}) and (\ref{eq7}). In reality, however,
this substitution radically changes the asymptotic behavior of the free energy 
at low temperature:
\begin{eqnarray}
&&
\tilde{\cal F}(a,T)={\cal F}(a,T)-\frac{k_BT}{16\pi a^2}
\left[\zeta(3)-{\rm Li}_3(r_0^2)+R(T)\right],
\nonumber \\
&&
\tilde{\cal F}^{A}(a,T)={\cal F}^{A}(a,T)-\frac{k_BT}{4 a^3}
\left[(1-r_0)\alpha(0)+\tilde{R}(T)\right].
\label{eq12}
\end{eqnarray}
\noindent
Here, ${\rm Li}_n(z)$ is the polylogarithm function, ${\cal F}(a,T)$ and
${\cal F}^{A}(a,T)$ are given by (\ref{eq6}) and (\ref{eq7}), respectively,
and $R(T),\>\tilde{R}(T)$ are exponentially vanishing functions 
when $T\to 0$.
In this case the definition of the Casimir entropy (\ref{eq8})
results in the following nonzero values at zero temperature [5,20]
\begin{eqnarray}
&&
\tilde{S}(a,0)=\frac{k_B}{16\pi a^2}
\left[\zeta(3)-{\rm Li}_3(r_0^2)\right]>0,
\nonumber \\
&&
\tilde{S}^{A}(a,0)=\frac{k_B}{4 a^3}
(1-r_0)\alpha(0)>0.
\label{eq13}
\end{eqnarray}
\noindent
This is in violation of the third law of thermodynamics (the Nernst
theorem) because not only are these values not equal to zero, but they
also depend on the parameters of the system (separation distance, static
dielectric permittivity and atomic polarizability).

The obtained result can be considered as a puzzle because instead of more
accurate asymptotic expressions one arrives to the conclusion of an
apparent contradiction of the Lifshitz theory with the fundamental principle
of thermodynamics.

\section{Experimental tests}

Rapid progress in the measurements of the Casimir force is very helpful
for the resolution of the theoretical controversies. As mentioned in the
Introduction, the experiment [6,7] excluded the use of the Drude model to
describe the thermal Casimir force between metal plates. It is important
to note, however, that a much smaller thermal effect, as predicted by
the generalized plasma model [24,25], was not registered in that
experiment. Here we consider the results of two other experiments involving
semiconductor and dielectric materials [14--16]. Both of them are 
accurate enough to resolve between the theoretical approaches including 
and neglecting the dielectric dc conductivity. What is more,
one of them [16] provides the first experimental observation of the
thermal effects in Casimir physics.

We start with the observation of modulation of the Casimir force between
a gold coated sphere and a Si plate with laser light [14,15]. In this
experiment, a B doped $p$-type Si plate was illuminated with laser pulses
and the difference of the Casimir forces between a sphere and a plate
in the presence and in the absence of light has been measured.
In the absence of laser pulse the concentration of charge carriers in the
Si plate was $\tilde{n}=5\times 10^{14}\,\mbox{cm}^{-3}$ and in the presence 
of light $n_1=2.1\times 10^{19}\,\mbox{cm}^{-3}$ or
$n_2=1.4\times 10^{19}\,\mbox{cm}^{-3}$ for the two different absorbed
powers $P_{w1}=9.3\,$mW and $P_{w2}=4.7\,$mW.
The directly measured quantity was, thus,
$\Delta F(a)=F^{L}(a)-F(a)$, where $F^{L}$ and $F$ are the Casimir forces
in the presence and in the absence of laser light on the plate,
respectively. The experimental data were compared with the Lifshitz theory
either neglecting or including dc conductivity of the high-resistivity Si
in the dark phase.

In Fig.~1(a,b) the mean experimental data $\langle\Delta F(a)\rangle$
are shown as dots for the absorbed powers $P_{w1}$ and $P_{w2}$, respectively. 
The calculational results for $\Delta F(a)$, using the Lifshitz theory
with neglect of the dc conductivity of the high-resistivity Si (dark phase),
are shown as the solid lines. Similar calculational results using the
Lifshitz theory with included dc conductivity (\ref{eq11}) in the
dark phase are shown as the long-dashed lines. As is seen in Fig.~1(a,b)
the solid lines are consistent with the data whereas the 
long-dashed lines are 
excluded over the wide separation region. In the same figure, the 
short-dashed
lines show the computational results obtained from the Lifshitz formula
at zero temperature. As is seen in the figure, the computational
results at $T=0$ are consistent with the data. These results do not
depend on whether the dc conductivity was included or neglected.
The experiment under consideration is precise enough to exclude the large
thermal effect, as obtained when the dc conductivity in the dark phase
is included, but is not of sufficient precision to measure the small
thermal effect obtained in the case when the dc conductivity in the dark 
phase is neglected.
\begin{figure}[t]
\begin{center}
\vspace*{-16cm}
\hspace*{-5.cm}
\includegraphics[width=25cm]{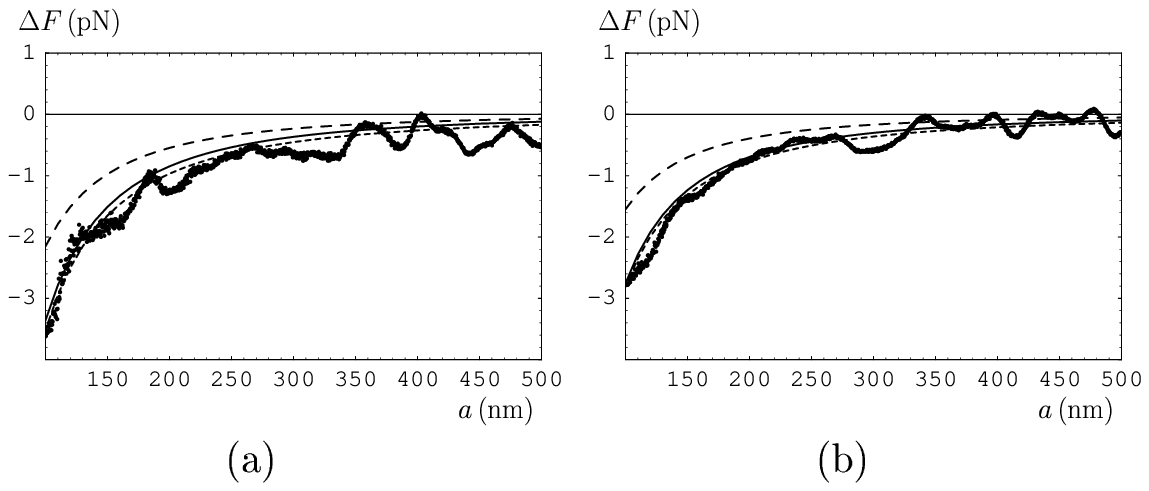}
\vspace*{-15.2cm}
\end{center}\caption{Differences of the Casimir forces in the
presence and in the absence of light versus separation for different
absorbed powers: (a) 9.3\,mW and (b) 4.7\,mW. The mean measured differences
 are shown as dots. Differences calculated using the Lifshitz
formula at $T=300\,$K and $T=0$ are shown as the solid and short-dashed
lines, respectively, and those calculated including the dc conductivity of
high-resistivity Si as the long-dashed lines.} 
\end{figure}

For illustrative purpose, in Fig.~2 we demonstrate the comparison between 
experiment and theory over a more narrow separation interval from 100 to 
150\,nm where only each third experimental point is plotted together with
its error bars $[\pm\Delta a,\,\pm\Delta^{\!\rm tot}(\Delta F^{\rm expt})]$
shown as crosses (the absorbed power is 8.5\,mW).
\begin{figure}[t]
\begin{center}
\vspace*{-9cm}
\hspace*{-4.cm}
\includegraphics{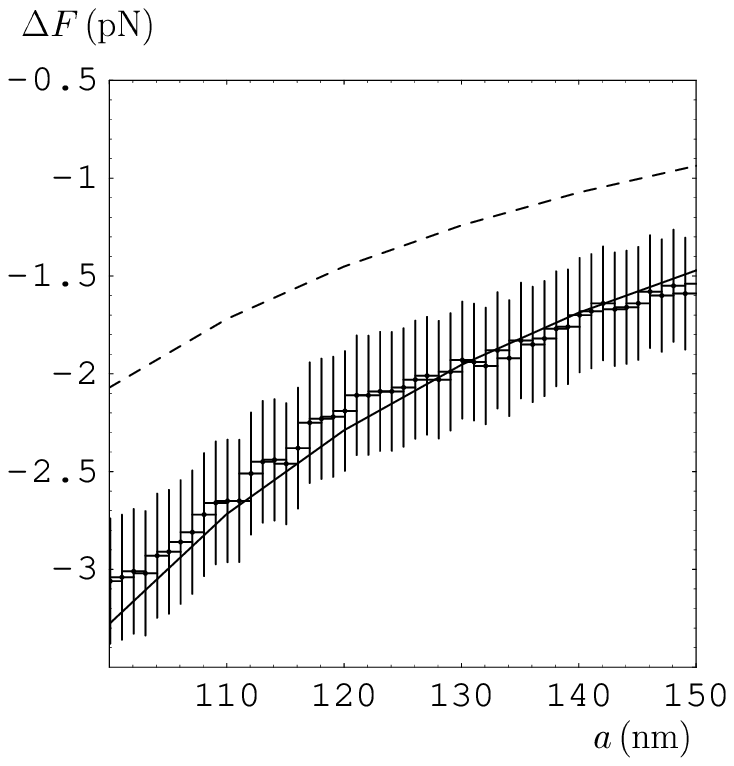}
\vspace*{-14cm}
\end{center}\caption{Mean experimental differences of the Casimir
forces with their experimental errors are shown as crosses. 
Solid and dashed lines represent the theoretical differences 
computed at $T=300\,$K using the model of dielectric permittivity
of high-resistivity Si neglecting and including the dc conductivity,
respectively.} 
\end{figure}
Note that these errors are entirely experimental. They are determined at a
95\% confidence level as a combination of random and systematic errors
and are not connected with any theory.
The theoretical force differences $\Delta F^{\rm theor}$ computed at
$T=300\,$K with neglected and included dc conductivity in the dark
phase are shown as the solid and dashed lines, respectively.
{}From Fig.~2 one concludes that inclusion of the dc conductivity in the
dark phase is excluded by the experimental data at a 95\% confidence
level within the separation region from 100 to 150\,nm.

Note that computations of the Casimir force between an Au coated sphere
and Si plate were done using the proximity force approximation [26]
\begin{equation}
F(a,T)=2\pi R{\cal F}(a,T),
\label{eq14}
\end{equation}
\noindent
where ${\cal F}(a,T)$ is given by (\ref{eq1}). The dielectric permittivities
$\varepsilon^{(1)}({\rm i}\xi)$ and $\varepsilon^{(2)}({\rm i}\xi)$
for Si and Au, respectively, were obtained from the tabulated optical
data [23] by means of the Kramers-Kronig relations. Practically the same
results can be obtained using the analytical representations for
$\varepsilon^{(n)}({\rm i}\xi)$. For Au high precision six-oscillator
model is considered in [7]. For Si, the imaginary part of the dielectric
permittivity can be modeled as a constant within some frequency region [27]
\begin{equation}
{\rm Im}\,\varepsilon^{(1)}(\omega)=\left\{
\begin{array}{lll}
\bar{\varepsilon}, &{\ \ } & \omega_0\leq\omega\leq\omega_1, \\
0, && \omega<\omega_0,{\ }\omega>\omega_1,
\end{array}\right.
\label{eq15}
\end{equation}
\noindent
where $\omega_0=3.22\,$eV, $\omega_1=4.62\,$eV, $\bar{\varepsilon}=48$.
Using the Kramers-Kronig relation, this results in
\begin{equation}
\varepsilon^{(1)}({\rm i}\xi)=1+\frac{\bar{\varepsilon}}{\pi}\,
\ln\left(\frac{\omega_1^2+\xi^2}{\omega_0^2+\xi^2}\right).
\label{eq16}
\end{equation}
\noindent
The use of the approximation (\ref{eq16}) instead of the tabulated optical
data [23] leads to less than 1\% deviations in the Casimir force (\ref{eq14})
between an Au sphere and Si plate. At separations $a=50$, 100, 200, 400
and 600\,nm this deviation is equal to $-0.58$\%, 0.30\%, 0.70\%, 0.84\%
and 0.86\%, respectively.

Another analytic formula representing the dielectric permittivity
of Si along the imaginary frequency axis takes the form [28]
\begin{equation}
\varepsilon^{(1)}({\rm i}\xi)=\varepsilon^{(1)}({\rm i}\infty)
+\frac{\bigl[\varepsilon^{(1)}({\rm i}0)-
\varepsilon^{(1)}({\rm i}\infty)\bigr]\omega_0^2}{\omega_0^2+\xi^2},
\label{eq17}
\end{equation}
\noindent
where $\omega_0=6.6\times 10^{15}\,$rad/s, 
$\varepsilon^{(1)}({\rm i}0)=11.87$ and
$\varepsilon^{(1)}({\rm i}\infty)=1.035$.
This representation also leads to less than 1\% differences in the case 
of Au sphere above Si plate in comparison to computations using the
tabulated optical data for Si. Thus, at separations $a=50$, 100, 200, 400
and 600\,nm the relative deviations between both computations are
equal to 0.75\%, 0.74\%, 0.62\%, 0.55\% and 0.52{\%}.
Because of this, both the tabulated optical data for Si and different
analytical representations for its dielectric permittivity can be used
in the computations within 1\% precision.

Another important experimental test is the measurement of the thermal
Casimir-Polder force through center-of-mass oscillations
of the trapped Bose-Einstein condensate [16]. In this experiment, not
only were different models of the dielectric response compared with the
data [16,29], but the thermal effect was measured for the first time in
Casimir physics. The condensate of Rb atoms was at a distance from 7 to 
$11\,\mu$m from a fused SiO${}_2$ dielectric substrate. In the original
publication this substrate was considered as a dielectric material with
the static dielectric permittivity $\varepsilon^{(1)}(0)=3.81$.
However, at room temperature (in an equilibrium situation the temperature
of the substrate and environment were equal to $T_S=T_E=310\,$K)
SiO${}_2$ has nonzero conductivity which is ionic in nature and varies
from $10^{-9}\,\mbox{s}^{-1}$ to $10^{2}\,\mbox{s}^{-1}$ depending on
the concentration of alkali ions which are always present as trace
constituents [30,31]. This conductivity was taken into account in [29] in 
accordance with (\ref{eq11}).

The directly measured quantity in the experiment [16] was the relative
shift of the frequency of center-of-mass oscillations of the 
Bose-Einstein condensate due to the presence of the Casimir-Polder
force
\begin{equation}
\gamma_z=\frac{|\omega_0-\omega_z|}{\omega_0}=
\frac{1}{2m\omega_0^2}\,|\langle\partial_zF^{A}(z,T)\rangle | .
\label{eq18}
\end{equation}
\noindent
Here, $m=1.443\times 10^{-25}\,$kg is the mass of the Rb atom,
$\omega_0$ is the unperturbed trap frequency 
$\omega_0=2\pi\times 229\,$rad/s, $F^{A}(z,T)$ in the equilibrium situation
is given by the negative derivative of ${\cal F}^{A}(z,T)$ defined in
(\ref{eq3}) with respect to $z$, and the averaging is done over the period
of condensate oscillations and over the volume of the condensate cloud.

\begin{figure}[b]
\begin{center}
\vspace*{-12.5cm}
\hspace*{-2.cm}
\includegraphics{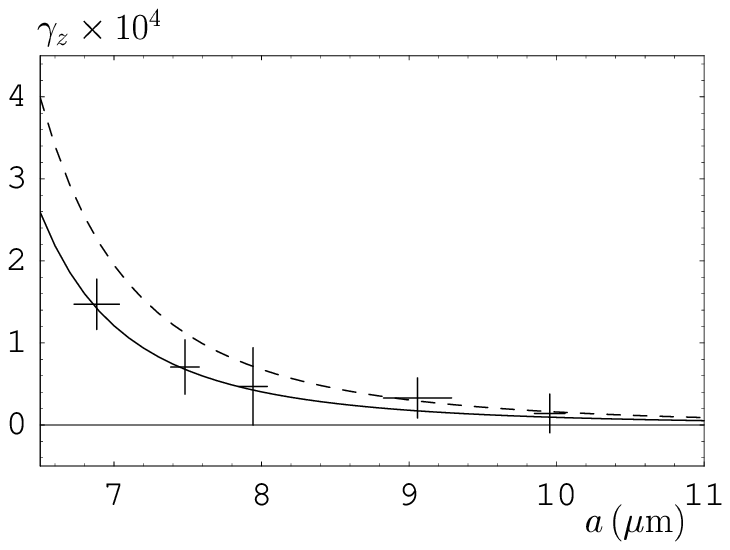}
\vspace*{-12.5cm}
\end{center}\caption{
Fractional change in the trap frequency versus separation in thermal
equilibrium with $T_S=T_E=310\,$K computed by neglecting (solid line)
and including  (dashed line) the conductivity of the dielectric
substrate. The experimental data are shown as crosses.} 
\end{figure}
The experimental data for the relative frequency shift $\gamma_z$ are 
presented as crosses in Fig.~3. The arms of the crosses show the total 
experimental errors determined in [16] at a 67\% confidence level
individually for each data point. The results of theoretical computations
using (\ref{eq18}) with neglected dc conductivity of SiO${}_2$
substrate are shown by the solid line [16,29]. As is seen in the figure,
theory neglecting the dc conductivity of fused silica is in a very good
agreement with the data. In the same figure, the results of theoretical
computations by (\ref{eq18}) with dc conductivity of SiO${}_2$ included
are shown by the dashed line [29]. As is seen in the figure, the dashed 
line is in evident disagreement with the first two 
experimental points.

The experiment was repeated for two more times in situations out of
thermal equilibrium ($T_S=479\,$K, $T_E=310\,$K and $T_S=605\,$K, 
$T_E=310\,$K), i.e., with the temperature of the plate higher than the
environment temperature. In this case the Casimir-Polder force $F^{A}(z,T)$
in (\ref{eq18}) should be replaced with
\begin{equation}
F^A(z,T_S,T_E)=F^A(z,T_E)+F^{\rm neq}(z,T_S)-F^{\rm neq}(z,T_E),
\label{eq19}
\end{equation}
\noindent where the explicit form of the function $F^{\rm neq}(z,T)$ can be
found in [16,29,32].

\begin{figure}[t]
\begin{center}
\vspace*{-15cm}
\hspace*{-1.5cm}
\includegraphics{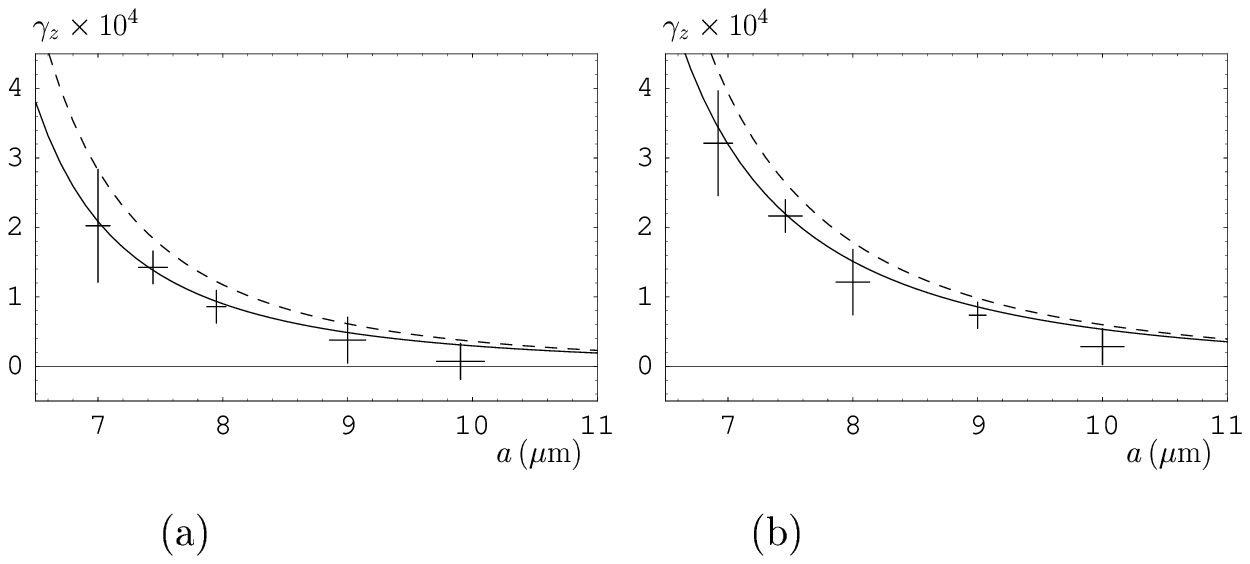}
\vspace*{-11.cm}
\end{center}\caption{
Fractional change in the trap frequency versus separation
out of  thermal equilibrium with 
(a) $T_S=479\,$K, $T_E=310\,$K
and (b) $T_S=605\,$K, $T_E=310\,$K.
Computations are done by neglecting (solid line)
and including (dashed line) the conductivity of the dielectric
substrate. The experimental data are shown as crosses.} 
\end{figure}
In Fig.~4 the experimental data for $\gamma_z$ are shown for 
(a) $T_S=479\,$K  and (b) $T_S=605\,$K as crosses with their arms
indicating the experimental
errors. As in Fig.~3, the theoretical results computed with the dc
conductivity of SiO${}_2$ plate neglected and included are shown by
the solid and dashed lines, respectively. As is seen in Fig.~4(a),
the dashed line is in disagreement with 
the three experimental points,
and the other two only touch it. In Fig.~4(b)  all data points exclude 
the theoretical approach represented by the dashed line.
Thus, the inclusion of the dc conductivity of the dielectric substrate
into the Lifshitz theory leads to disagreement with the experimental
data. The calculational results using the Lifshitz theory with the dc
conductivity neglected are in very good agreement with the experimental
data. We emphasize that the theoretical results obtained with dc
conductivity included do not depend on the value of conductivity, but only
of the fact that it is nonzero. Importantly, the inclusion of
conductivity does not influence the nonequilibrium contributions to
$\gamma_z$. Thus, the puzzling effect comes only from the contribution
to $\gamma_z$ given by the Lifshitz theory.

\section{Phenomenological approach and its justification}

In the above, both theoretical and experimental problems are elucidated 
arising in the Lifshitz theory when applied to dielectric materials at
nonzero temperature with dc conductivitincludedy. Recently, these problems
were discussed also in [33] but no solution was found. In [33]
only ``weakly conducting'' materials were considered which possess nonzero 
conductivity at zero temperature. These are in fact metallic-type
semiconductors with a dopant concentration above critical. To satisfy
the Nernst heat theorem, it was suggested that there is nonzero residual
relaxation at $T=0$, i.e., the crystal lattice is not perfect.
Materials of such type have nothing to do with dielectrics whose
conductivity vanishes when temperature goes to zero. The formalism of [33]
is also in contradiction with precision experiments dealing with
metallic test bodies [6,7].

To avoid contradictions of the Lifshitz theory with the principles of
thermodynamics and  experimental results, one can propose the
following phenomenological prescription. In the applications
of the Lifshitz theory to all materials having zero conductivity at zero
temperature, the presence of free charge carriers at nonzero
temperature should be disregarded. The dielectric permittivity
of such materials should be described by equation (\ref{eq4}).
In the applications of the Lifshitz theory to all materials having nonzero
conductivity at zero temperature, free charge carriers should be included
in the framework of the generalized plasma-like model
\begin{equation}
\varepsilon_{\rm gp}({\rm i}\xi)=1+\frac{\omega_p^2}{\xi^2}+\sum_{j}
\frac{g_j}{\omega_j^2+\xi^2+\gamma_j\xi},
\label{eq20}
\end{equation}
\noindent
where all $\omega_j\neq 0$. This prescription is consistent with
thermodynamics and all available experimental data [22,24,25,34].

The physical reasons of why the Lifshitz theory does 
not allow the
inclusion of real conductivity processes can be understood as follows.
Lifshitz derived his famous formulas under the condition of thermal
equilibrium. This means that not only $T={\rm const}$, but also all
irreversible processes connected with the dissipation of energy
into heat have already been terminated [35,36]. 
The Drude-like dielectric
function (\ref{eq11}) is derived from the Maxwell equations with a
real drift current of conduction electrons 
$\mbox{\boldmath$j$}=\sigma_0\mbox{\boldmath$E$}$ initiated by the
external electric field $\mbox{\boldmath$E$}$ [24].
The drift current is an irreversible process which
takes a system out of thermal equilibrium. This
 current leads to Joule heating of the Casimir plates
(Ohmic losses) [37].
To preserve the temperature constant, one should admit that there
exists an unidirectional
flux of heat from the medium to the heat reservoir [38]. 
Such interactions between
a system and a heat reservoir are prohibited by the definition of
 thermal equilibrium [39]. 
One can conclude that the substitution of the Drude-like dielectric
permittivities into the Lifshitz formulas violate the applicability
conditions under which this theory was developed. Then it is not
surprising that such substitution results in problems with thermodynamics and
contradictions with experimental data. {}From this point of view, the
Lifshitz theory is free of problems and paradoxes which arise only if it is
applied incorrectly.

Another question to be answered is the following. One could agree with the
fact that the conductivity processes in real materials connected with the
drift current of conduction electrons violate the applicability
conditions of the Lifshitz theory and, thus, are not described by this
theory. These processes, however, exist in reality. Because of this should
they be taken into account in some future general theory of dispersion
forces? At the moment we have no answer to this question.
All computations on the basis of the Lifshitz theory for dielectrics with
dc conductivity neglected 
and for metals by using the plasma-like dielectric permittivity
(\ref{eq20}) are in agreement with both thermodynamics and experimental data.
This suggests that the conductivity properties are not related to
dispersion forces. More precise experiments will probably bring
additional information to bear on this subject.

\section{Approaches based on the modification of reflection coefficients
due to screening effects}

Recently it was proposed [17,18] that the problems of the Lifshitz theory 
described above may have an alternative solution if
the screening effects due to free charge carriers are taken into account. 
This approach includes both the drift and diffusion currents
of free charge carriers through use of the transport Boltzmann equation [18].
It uses the standard Lifshitz formulas (\ref{eq1}) and 
(\ref{eq3}) for the free energy of wall-wall and atom-wall
interaction with the TE reflection coefficient, $\tilde{r}_{\rm TE}$,
as defined in (\ref{eq2}) within the Drude model approach, 
but with the modified TM reflection
coefficient
\begin{equation}
\tilde{r}_{\rm TM}({\rm i}\xi,k_{\bot})=
\frac{\tilde\varepsilon({\rm i}\xi)q-k-\frac{k_{\bot}^2}{\eta({\rm i}\xi)}\,
\frac{\tilde\varepsilon({\rm i}\xi)-
\varepsilon({\rm i}\xi)}{\varepsilon({\rm i}\xi)}}{\tilde\varepsilon
({\rm i}\xi)q
+k+\frac{k_{\bot}^2}{\eta({\rm i}\xi)}\,
\frac{\tilde\varepsilon({\rm i}\xi)-
\varepsilon({\rm i}\xi)}{\varepsilon({\rm i}\xi)}}.
\label{eq21}
\end{equation}
\noindent
Here, 
\begin{equation}
\tilde\varepsilon({\rm i}\xi)=\varepsilon({\rm i}\xi)+
\frac{\omega_p^2}{\xi(\xi+\gamma)}, \qquad
k^2=k_{\bot}^2+\tilde\varepsilon({\rm i}\xi)\frac{\xi^2}{c^2}, 
\label{eq22}
\end{equation}
\noindent
where $\varepsilon({\rm i}\xi)$ is the permittivity of the bound
core electrons given in (\ref{eq4}).
The quantity $\eta({\rm i}\xi)$ is defined as
\begin{equation}
\eta({\rm i}\xi)=\left[k_{\bot}^2+\kappa^2
\frac{\varepsilon(0)}{\varepsilon({\rm i}\xi)}\,
\frac{\tilde\varepsilon({\rm i}\xi)}{\tilde\varepsilon({\rm i}\xi)-
\varepsilon({\rm i}\xi)}\right]^{1/2},
\label{eq23}
\end{equation}
\noindent
where $1/\kappa$ is the screening length.
If the charge carriers of density $n$ are described by the classical
Maxwell-Boltzmann statistics, as assumed in [18],
one gets the Debye-H\"{u}ckel screening length [40,41]
\begin{equation}
R_D=\frac{1}{\kappa}=\left(\frac{\varepsilon_0 k_BT}{4\pi e^2n}\right)^{1/2}. 
\label{eq24}
\end{equation}
\noindent
Assuming 
Fermi-Dirac statistics, one arrives at the Thomas-Fermi 
screening length [40,41]
\begin{equation}
R_{TF}=\frac{1}{\kappa}=
\left(\frac{\varepsilon_0 E_F}{6\pi e^2n}\right)^{1/2}.
\label{eq24a}
\end{equation}
\noindent
Here, $e$ is the electron charge, 
$\varepsilon_0=\varepsilon(0)$ is the dielectric
constant due to the core electrons, and $E_F=\hbar\omega_p$ is the
Fermi energy.

Note that the paper [18] applies this
approach to intrinsic semiconductor media. Bearing in mind,
however, that the classical Drude dielectric permittivity is
exploited, it is justified to apply it for metals and doped 
semiconductors (both dielectric and metallic type)
as well, by substituting the respective screening length
defined by (\ref{eq24}) or (\ref{eq24a}).

Now we consider whether or not the approach taking the screening effects
into account is consistent with thermodynamics. We concentrate our attention 
on the case of dielectric materials whose conductivity goes to zero
when the temperature vanishes. We start from the case of two similar
dielectric semispaces 
[$\varepsilon^{(1)}=\varepsilon^{(2)}=\tilde\varepsilon$].
The Casimir free energy per unit area is given by (\ref{eq1})  with
the reflection coefficients
$r_{\rm TE}^{(1)}=r_{\rm TE}^{(2)}=\tilde{r}_{\rm TE}$ 
from (\ref{eq2}), (\ref{eq22}), and 
$r_{\rm TM}^{(1)}=r_{\rm TM}^{(2)}=\tilde{r}_{\rm TM}$ 
from (\ref{eq21}). 

The derivation of the asymptotic behavior of the Casimir free energy
and entropy at low temperature is performed in perfect analogy with the
derivation of (\ref{eq6}) and (\ref{eq12}) in [5]. First we introduce the
small parameter ($\xi\geq\xi_1$)
\begin{equation}
\beta=\frac{4\pi\sigma(T)}{\xi},
\label{eq24b}
\end{equation}
\noindent
which goes to zero due to (\ref{eq10}) when the temperature vanishes.
Then we expand the reflection coefficients
$\tilde{r}_{\rm TM,TE}$ at all nonzero Matsubara frequencies
in powers of the parameter $\beta$
\begin{eqnarray}
&&
\tilde{r}_{\rm TM}({\rm i}\zeta,y)={r}_{\rm TM}({\rm i}\zeta,y)+
\beta\frac{\varepsilon y[2y^2+(\varepsilon-2)\zeta^2]}{\sqrt{y^2+
(\varepsilon-1)\zeta^2}[\varepsilon y+\sqrt{y^2+
(\varepsilon-1)\zeta^2}]^2} +O(\beta^2),
\nonumber \\
&&
\tilde{r}_{\rm TE}({\rm i}\zeta,y)={r}_{\rm TE}({\rm i}\zeta,y)+
\beta\frac{y[y-\sqrt{y^2+(\varepsilon-1)\zeta^2}]}{\sqrt{y^2+
(\varepsilon-1)\zeta^2}[ y+\sqrt{y^2+
(\varepsilon-1)\zeta^2}]} +O(\beta^2).
\label{eq25}
\end{eqnarray}
\noindent
Here, the reflection coefficients ${r}_{\rm TM,TE}$ are defined by (\ref{eq2})
with the dielectric permittivity (\ref{eq4}).
They lead to the asymptotic behavior of the Casimir free energy per unit
area given in (\ref{eq6}). The dimensionless variables $\zeta$ and $y$
are connected with other dimensional ones by the equations
\begin{equation}
y=2aq,\qquad \zeta=\frac{\xi}{\omega_c}=\frac{2a\xi}{c},
\label{eq26}
\end{equation}
\noindent
where $q=q(\xi)$ is defined in (\ref{eq2a}).

Now we substitute (\ref{eq25}) in (\ref{eq1}) and arrive at the following
expression for the Casimir free energy per unit area taking the screening
effects into account
\begin{equation}
\tilde{\cal F}(a,T)={\cal F}(a,T)+\frac{k_BT}{16\pi a^2}\left\{
\int_{0}^{\infty}\!\!\!y\,dy\ln\left[1-\tilde{r}_0^2(y){\rm e}^{-y}\right]
+{\rm Li}_3(r_0^2)+Q(T)
\right\},
\label{eq27}
\end{equation}
\noindent
where
\begin{equation}
\tilde{r}_0(y)\equiv\frac{\varepsilon_0\sqrt{y^2+(2a\kappa)^2}-
y}{\varepsilon_0\sqrt{y^2+(2a\kappa)^2}+y}
\label{eq28}
\end{equation}
\noindent
and $Q(T)$ vanishes exponentially when $T\to 0$.
The free energy ${\cal F}(a,T)$ in (\ref{eq27}) is given by (\ref{eq6}),
and $\kappa$  in (\ref{eq28}) is defined in (\ref{eq24}).
Calculating the negative derivative of both sides of (\ref{eq27})
with respect to $T$, 
we obtain the asymptotic behavior of the Casimir entropy at low
temperature
\begin{eqnarray}
&&
\tilde{S}(a,T)=S(a,T)-\frac{k_B}{16\pi a^2}\left\{
\vphantom{\int\limits_{0}^{0}}
\int_{0}^{\infty}\!\!\!y\,dy\ln\left[1-\tilde{r}_0^2(y){\rm e}^{-y}\right]
+{\rm Li}_3(r_0^2)\right.
\label{eq29} \\
&&~~~\left.
-8a^2\varepsilon_0T\frac{\partial\kappa^2}{\partial T}
\int_{0}^{\infty}\!\!\!dy\frac{y^2\tilde{r}_0(y)}{{\rm e}^{y}-
\tilde{r}_0^2(y)}\,\frac{1}{\sqrt{y^2+(2a\kappa)^2}[\varepsilon_0
\sqrt{y^2+(2a\kappa)^2}+y]^2}+Q(T)+TQ'(T)
\vphantom{\int\limits_{\infty}^{\infty}}\right\},
\nonumber
\end{eqnarray}
\noindent
where $S(a,T)$ is defined in (\ref{eq9}). It can be easily seen that the last
three terms in the curly brackets on the right-hand side of this
equation go to zero when $T$ goes to zero for any dielectric material.

The behavior of the first two terms in the curly brackets on the right-hand
side of (\ref{eq29}) when $T$ goes to zero is more involved.
If $n(T)$ exponentially decays to zero with vanishing temperature
(as is true for pure insulators and intrinsic semiconductors), then
according to (\ref{eq24}) so does $\kappa$. As a result,
$\tilde{r}_0(y)\to r_0=(\varepsilon_0-1)/(\varepsilon_0+1)$ and the
first two terms in the curly brackets cancel. Then the Casimir entropy
$\tilde{S}(a,T)$ goes to zero when $T$ vanishes following the same law
as $S(a,T)$, i.e., as $T^2$. This means that for insulators and intrinsic 
semiconductors the formalism under consideration is in agreement with the 
Nernst heat theorem.

However, there is a wide class of dielectric materials (such as doped
semiconductors with dopant concentration below critical and solids with
ionic conductivity) for which $n$ does not go to zero when $T$ goes to zero.
In fact, conductivity can be presented as [41]
\begin{equation}
\sigma(T)=n\,|e|\,\mu,
\label{eq30}
\end{equation}
\noindent
where $\mu$ is the mobility of charge carriers. Although $\sigma(T)$
goes to zero exponentially fast for all dielectrics when $T$ goes to
zero, for most of them this happens due to the vanishing mobility.
For instance, as mentioned in Sec.~3, the conductivity of SiO${}_2$ used
as the plate material in the experiment [16] is ionic in nature and is
determined by the concentration of impuritues. For all such materials,
in accordance with (\ref{eq24}), $\kappa\to\infty$ when $T\to 0$.
As a result, $\tilde{r}_0(y)\to 1$ when $T$ goes to zero in accordance
with (\ref{eq28}). In this case we obtain from (\ref{eq29})
\begin{equation}
\tilde{S}(a,0)=\frac{k_B}{16\pi a^2}\left[\zeta(3)-
{\rm Li}_3(r_0^2)\right]>0
\label{eq31}
\end{equation}
\noindent
in violation of the Nernst heat theorem. This means that for a wide
class of dielectric materials the proposed approach taking the screening 
effects into account is thermodynamically inconsistent in the same way as
the standard Lifshitz theory with the dc conductivity included
[compare with (\ref{eq13})].

We emphasize that the existence of dielectric materials for which $n$
does not go to zero but $\mu$ does go to zero when $T$ vanishes
demonstrates that the reflection coefficient (\ref{eq21}) at $\xi=0$ is
ambiguous. In reality, for such materials 
$\tilde{r}_{\rm TM}(0,k_{\bot})\to 1$ when $T$ and $\mu$ simultaneously 
vanish. This is because $\kappa\to\infty$ when $T\to 0$ in disagreement with
physical intuition that there should be no screening at zero mobility.
This ambiguity is connected with the break of continuity of
the reflection coefficient $\tilde{r}_{\rm TM}({\rm i}\xi,k_{\bot})$
at the point $\xi=0$, $T=0$.
If one takes the limit $T\to 0$ first, keeping 
$\xi={\rm const}\neq 0$, the standard Fresnel reflection coefficients
${r}_{\rm TM}$ from (\ref{eq2})  with no screening is reproduced.
This property is preserved in the subsequent limiting
transition $\xi\to 0$. As was noted in [42] on general grounds,
the violation of the Nernst heat theorem is caused by the breaks
in the continuity of the reflection coefficients at the origin of the
$(\xi,T)$-plane.

The theoretical approach taking the screening effects into account can
also be applied to describe the Casimir-Polder atom-wall interaction.
In this case it is also inconsistent with thermodynamics for dielectric
materials with a nonvanishing density of charge carriers. To find
the asymptotic behavior of the free energy at low temperature, one should
substitute into (\ref{eq3}) the reflection coefficients
$r_{\rm TM,TE}^{(1)}=\tilde{r}_{\rm TM,TE}$, as defined in (\ref{eq25}).
The result is
\begin{equation}
\tilde{\cal F}^{A}(a,T)={\cal F}^{A}(a,T)-
\frac{k_BT\alpha(0)}{8 a^3}\left[
\int_{0}^{\infty}\!\!\!y^2\,dy\,\tilde{r}_0(y){\rm e}^{-y}
-2r_0
+Q_1(T)\right],
\label{eq32}
\end{equation}
\noindent
where the asymptotic behavior of
 ${\cal F}^{A}(a,T)$  is presented in (\ref{eq7}), and $Q_1(T)$ vanishes
exponentially when $T\to 0$.
Calculating the negative derivative of both sides of (\ref{eq32})
with respect to $T$, one obtains
\begin{eqnarray}
&&
\tilde{S}^{A}(a,T)=S^{A}(a,T)+\frac{k_B\alpha(0)}{8 a^3}\left[
\vphantom{\int\limits_{0}^{0}}
\int_{0}^{\infty}\!\!\!y^2\,dy\,\tilde{r}_0(y){\rm e}^{-y}-2r_0\right.
\label{eq33} \\
&&~~~\left.
+4a^2\varepsilon_0T\frac{\partial\kappa^2}{\partial T}
\int_{0}^{\infty}\!\!\!dy\frac{y^3{\rm e}^{-y}}
{\sqrt{y^2+(2a\kappa)^2}(\varepsilon_0
\sqrt{y^2+(2a\kappa)^2}+y)^2}+Q_1(T)+TQ_1'(T)
\vphantom{\int\limits_{0}^{0}}\right],
\nonumber
\end{eqnarray}
\noindent
where $S^{A}(a,T)$ is defined in (\ref{eq9}). For dielectric materials 
with exponentially decaying $n(T)$ in the limit of low $T$,
$\tilde{r}_0(y)\to r_0$ and all terms on the right-hand side of (\ref{eq33})
added to $S^{A}(a,T)$ go to zero when $T\to 0$. As a result,
$\tilde{S}^{A}(a,T)$ goes to zero as $T^3$ and the Nernst theorem is 
satisfied. On the contrary, for all dielectric materials with
nonvanishing $n$ we get $\tilde{r}_0(y)\to 1$ when $T\to 0$ and in this
case (\ref{eq33}) results in
\begin{equation}
\tilde{S}^{A}(a,0)=\frac{k_B\alpha(0)}{4a^3}(1-r_0)>0.
\label{eq34}
\end{equation}
\noindent
This means that the Nernst theorem is violated in the same way as in
(\ref{eq13}) for the standard Lifshitz theory with the dc conductivity
included.

\begin{figure}[t]
\begin{center}
\vspace*{-5.5cm}
\hspace*{-1.5cm}
\includegraphics{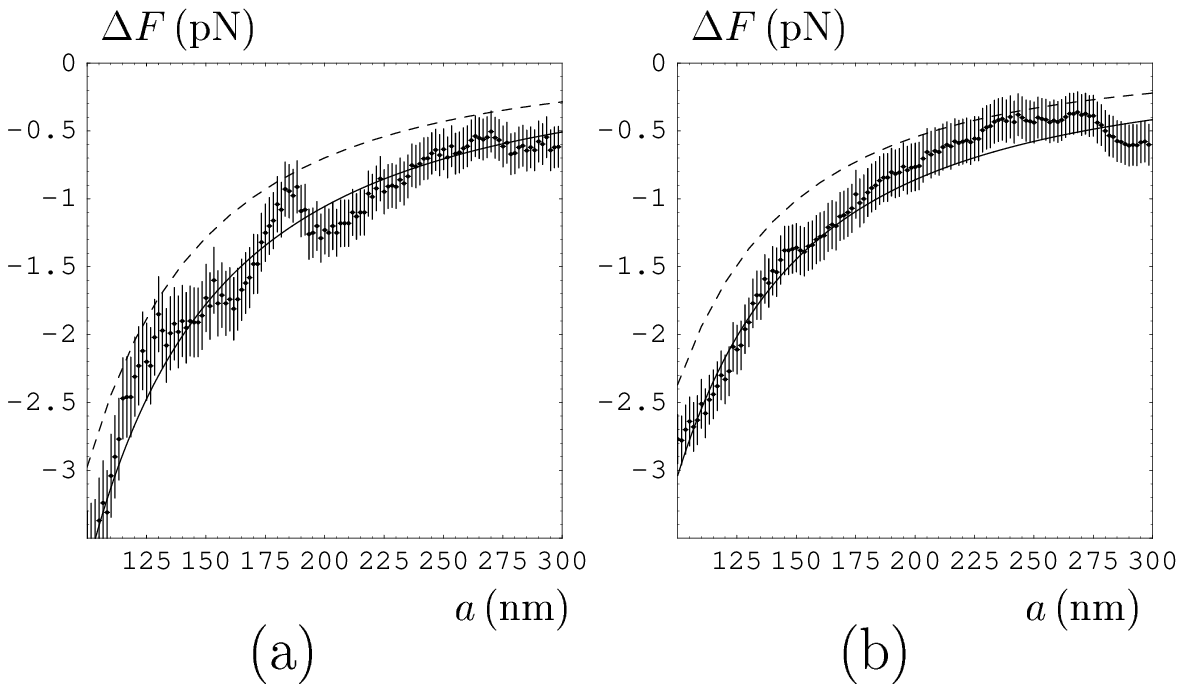}
\vspace*{-18.cm}
\end{center}\caption{
Differences of the Casimir forces between an Au-coated sphere
and a Si plate, in the presence and in the absence of laser light on
the plate, versus separation for the absorbed power of (a) 9.3\,mW and (b)
4.7\,mW.  The experimental data are shown as crosses.
Solid and dashed lines indicate the theoretical results calculated using
the standard Lifshitz theory with the dc conductivity of Si in the
dark phase neglected, and the generalization of this theory [17,18],
respectively.} 
\end{figure}
The theoretical predictions following from the approach proposed in [18] can
be compared with the measurement data of the experiment on the modulation of
the Casimir force with light (see Sec.~3 for details). In Fig.~5(a) and 
5(b) we present the experimental data of this experiment obtained for
the absorbed powers $P_{w1}$ and $P_{w2}$, respectively,
with the experimental errors of force measurements determined at 70\%
confidence level. The theoretical difference Casimir force, computed using 
the standard Lifshitz theory with the dc conductivity of
high-resistivity Si in the dark phase neglected,
is shown by the solid lines. 
The theoretical difference Casimir force computed by including the screening 
effects at all Matsubara frequencies, as described above, is shown by the 
dashed lines.
It can be seen that the standard Lifshitz theory is consistent with data, 
whereas the theoretical approach taking the screening effects into account
is excluded by the data at a 70\% confidence level.

In Sec.~4 the contradictions with thermodynamics and the experimental data
of the standard Lifshitz theory with the dc conductivity included are
explained based on the violation of its applicability condition, i.e., of
thermal equilibrium, in the presence of drift current. In a similar way,
the approach taking the screening effects into account applies the
Lifshitz theory to physical phenomena involving both the drift and diffusion
currents. The latter is caused by the nonequilibrium distribution of charge
carriers in an external field, i.e., by a physical situation out of
thermal equilibrium. This is because the diffusion current is
determined by a nonzero gradient of charge carrier density, whereas
for homogeneous systems in thermal equilibrium the charge carrier
density must be homogeneous. Thus, the problems and paradoxes arising
in this case are also explained by the use of the Lifshitz theory
outside of its defined application region.

An earlier attempt to incorporate the screening effects in the theory
of dispersion forces was undertaken in [17]. In this approach the effect
of screening was taken into account only for the static field in the
interaction of an atom with a dielectric wall. Bearing in mind that in
accordance with (\ref{eq3}) the TE reflection coefficient at zero
frequency does not contribute to the result, the single modified
coefficient was [17]
\begin{equation}
r_{\rm TM}^{\rm mod}(0,k_{\bot})=\frac{\varepsilon_0
\sqrt{k_{\bot}^2+\kappa^2}-k_{\bot}}{\varepsilon_0
\sqrt{k_{\bot}^2+\kappa^2}+k_{\bot}},
\label{eq35}
\end{equation}
\noindent
where $\kappa$ is given in (\ref{eq24}). It is easily seen that (\ref{eq35})
coincides with the general TM reflection coefficient
$\tilde{r}_{\rm TM}({\rm i}\xi,k_{\bot})$ defined in (\ref{eq21})
if one puts $\xi=0$. 

The reflection coefficient (\ref{eq35}) can be obtained 
by the formal replacement of the dielectric permittivities depending
only on the frequency with the permittivities depending both on frequency
and the wave vector in the Fresnel reflection coefficient
of an uniaxial crystal.
 To confirm that this is the case, one should introduce
nonisotropic permittivities 
$\varepsilon_x=\varepsilon_y\neq\varepsilon_z$ 
and replace the coefficients $r_{\rm TM}$ and
$r_{\rm TE}$ in  (\ref{eq2})
with
\begin{equation}
r_{\rm TM}^{\rm mod}({\rm i}\xi,k_{\bot})=
\frac{\sqrt{\varepsilon_x({\rm i}\xi)\varepsilon_z({\rm i}\xi)}q-
k_{z}}{\sqrt{\varepsilon_x({\rm i}\xi)\varepsilon_z({\rm i}\xi)}q+k_{z}},
\qquad
r_{\rm TE}^{\rm mod}({\rm i}\xi,k_{\bot})=
\frac{q-k_x}{q+k_x}.
\label{eq36}
\end{equation}
\noindent 
Here, $k_z$ and $k_x$ are defined by (\ref{eq2a}) with the replacement 
of $\varepsilon({\rm i}\xi)$ by $\varepsilon_z({\rm i}\xi)$ and
$\varepsilon_x({\rm i}\xi)$, respectively. For
$\varepsilon_x$ and $\varepsilon_z$ depending only on the frequency,
(\ref{eq36}) is commonly used for the description of uniaxial
crystals (see, e.g., [43,44]).
Then (\ref{eq35}) follows from (\ref{eq36}) if one puts
\begin{equation}
\varepsilon_x(0)=\varepsilon_0, \quad
\varepsilon_z(0)=\varepsilon_z(0,k_{\bot})=\varepsilon_0
\left(1+\frac{\kappa^2}{k_{\bot}^2}\right),
\label{eq37}
\end{equation}
\noindent 
where now $\varepsilon_z(0)$ depends on the wave vector, i.e., spatial
dispersion is present.
Thus, to obtain (\ref{eq35}) one substitutes into the Fresnel reflection
coefficient (\ref{eq36}) derived from the standard continuity boundary
conditions a dielectric permittivity depending on the wave vector
(see  [45,46]  why this is inappropriate).

We emphasize that the general reflection coefficients taking the
screening effects into account [the transverse electric from (\ref{eq2})
with the dielectric permittivity (\ref{eq22}) and the transverse magnetic
from (\ref{eq21})] cannot be represented in the form of (\ref{eq36}).
To prove this statement, we note that from the equality
$\tilde{r}_{\rm TE}={r}_{\rm TE}^{\rm mod}$ it follows that
$\tilde\varepsilon({\rm i}\xi)=\varepsilon_x({\rm i}\xi)$. Thus, in
accordance with (\ref{eq22}), $\varepsilon_x({\rm i}\xi)\to\infty$ when
$\xi\to 0$. Keeping in mind that $\varepsilon_z(0)>0$,
from (\ref{eq36}) one obtains $r_{\rm TM}^{\rm mod}(0,k_{\bot})=1$.
This is in contradiction with the above result that at zero frequency
the reflection coefficient (\ref{eq21}) coincides with the reflection
coefficient (\ref{eq35}). This means that the general approach taking
the screening effects into account [18] cannot be described in terms
of any dielectric permittivity which includes spatial dispersion.

The approach based on the modification of only one reflection coefficient
(\ref{eq35}) meets the same thermodynamical difficulties as the general
approach considered above. Thus, if one computes the Casimir entropy
with the reflection coefficient (\ref{eq35}) and all the other standard
ones, as given by (\ref{eq2}), it goes to zero when the temperature vanishes
for dielectrics with exponentially vanishing charge carrier density $n$.
However, for dielectrics with nonvanishing $n$ the result (\ref{eq34})
is reproduced, i.e., the Nernst heat theorem is violated [20].

The modification of the TM reflection coefficient in accordance with
(\ref{eq35}) can be verified experimentally by using the measurement
data of the optical modulation experiment discussed in Sec.~3.
In theoretical computations at zero frequency the expression (\ref{eq35})
was used. At all nonzero Matsubara frequencies the standard reflection
coefficients (\ref{eq2}) were substituted into the Lifshitz formula.
The computational results using the standard
reflection coefficients at all nonzero Matsubara frequencies coincide 
with those obtained with the general reflection coefficients
$\tilde{r}_{\rm TM}({\rm i}\xi,k_{\bot})$ and
$\tilde{r}_{\rm TM}({\rm i}\xi,k_{\bot})$.
Then we return to the dashed lines in Fig.~5(a,b) representing the
difference Casimir force computed with the inclusion of screening effects.
More generally, it can be easily seen that for materials with sufficiently
small charge carrier density (for instance for Si with
$n\leq 10^{17}\,\mbox{cm}^{-3}$) the general reflection coefficient
(\ref{eq21}) including the screening effects leads to
approximately the same values  at room temperature,
as the standard one at all nonzero
Matsubara frequencies. Thus, the theoretical results computed by using
the modified TM reflection coefficient (\ref{eq35}) are excluded by data
at a 70\% confidence level.

\section{Conclusions and discussion}

In the above we have considered the application of the Lifshitz
theory to real metals,
dielectrics and semiconductors. When the conductivity processes,
connected with the presence of free charge carriers, are included into the
model of the dielectric response, problems and paradoxes in the 
application of the Lifshitz
theory are immediately apparent. They are manifested in the
contradictions of the Lifshitz theory with both thermodynamics and
experiment. We have listed several measurements performed by different
experimental groups with metals, dielectrics and semiconductors which
are in contradiction with the theoretical results taking into account
the drift current of conduction electrons.

In searching for reasons of why the problems and paradoxes arise, it was
emphasized that the Lifshitz theory was formulated for systems in thermal 
equilibrium. As a result, this theory is not applicable to physical
systems where processes go on which violate thermal equilibrium.
Specifically, in thermal equilibrium all processes of heat transfer,
diffusion and others connected with the dissipation of energy are
terminated. A question then arises if the drift current of charge
carriers which leads to a unidirectional flux of Joule heat from the
Casimir plates to the heat reservoir is compatible with the state of
thermal equilibrium. Our conclusion is that it is not compatible with
thermal equilibrium and, thus, cannot be considered in the framework
of the Lifshitz theory. The inclusion of conductivity properties in the
Lifshitz theory is in violation of its basic applicability condition.

We have analyzed recent attempts to solve problems and
paradoxes using screening effects and diffusion processes. 
We have shown that these attempts are in conflict with thermodynamics
and in contradiction with experiment. This is because diffusion
processes make the deviations from thermal equilibrium even more
dramatic.

One may conclude that the Lifshitz theory, if used within its application
region, is free of problems and paradoxes. Problems and paradoxes arise
when the basic applicability condition of the Lifshitz theory,
thermal equilibrium, is not taken into account by using the Drude
model or introducing the diffusion current. 
What is more, phenomenologically the Lifshitz theory leads to both 
thermodynamically and experimentally consistent results if the processes
of electric conductivity connected with both drift and diffusion currents
are simply disregarded. Future progress should show if these processes have
any relation to the phenomenon of dispersion forces.

\ack{This work was supported by Deutsche Forschungsgemeinschaft,
 Grant No 436 RUS 113/789/0--4.
The author is grateful to the Center of Theoretical Studies and Institute
of Theoretical Physics of Leipzig University, where this work was 
performed, for their kind  hospitality.}

\smallskip

\end{document}